# Microstructural investigation of hybrid CAD/CAM restorative dental materials by micro-CT and SEM

Elisabeth Prause*[a], Jeremias Hey[b], Florian Beuer[a], Jamila Yassine[a], Bernhard Hesse[c], Timm Weitkamp[d], Javier Gerber[c], Franziska Schmidt[a]

[a] Department of Prosthodontics, Geriatric Dentistry and Craniomandibular Disorders, Charité-Universitätsmedizin Berlin, corporate member of Freie Universität Berlin and Humboldt-Universität zu Berlin, Aßmannshauser Str. 4-6, 14197 Berlin, Germany

[b] Department of Prosthodontics, School of Dental Medicine, Martin-Luther-University, Halle, Germany

[c] Xploraytion GmbH, Bismarckstr. 10-12, 10625 Berlin, Germany

[d] Synchrotron SOLEIL, L'Orme des Merisiers, Départementale 128, 91190 Saint-Aubin, France

*Corresponding author: Dr. Elisabeth Prause, Charité Universitätsmedizin Berlin, Institut für Zahn-, Mund- und Kieferheilkunde, Abteilung für Zahnärztliche Prothetik, Alterszahnmedizin und Funktionslehre, Aßmannshauser Str. 4-6, 14197 Berlin, Germany. Email: elisabeth.prause@charite.de

**Abstract**

Objectives: An increasing number of CAD/CAM (computer-aided design/computer-aided manufacturing) hybrid materials have been introduced to the dental market in recent years. In addition, CAD/CAM hybrid materials for additive manufacturing (AM) are becoming more attractive in digital dentistry. Studies on material microstructures using micro-computed tomography (µ-CT) combined with scanning electron microscopy (SEM) have only been available to a limited extent so far.

Methods: One CAD/CAM three-dimensional- (3D-) printable hybrid material (VarseoSmile Crown plus) and two CAD/CAM millable hybrid materials (Vita Enamic; Voco Grandio), as well as one direct composite material (Ceram.x duo), were included in the present study. Cylindrical samples with a diameter of 2 mm were produced from each material and investigated by means of synchrotron radiation µ-CT at a voxel size of 0.65 µm. Different samples from the same materials, obtained by cutting and polishing, were investigated by SEM.

Results: The 3D-printed hybrid material showed some agglomerations and a more irregular distribution of fillers, as well as a visible layered macrostructure and a few spherical pores due to the printing process. The CAD/CAM millable hybrid materials revealed a more homogenous distribution of ceramic particles. The direct composite material showed multiple air bubbles and microstructural irregularities based on manual processing.

Significance: The µ-CT and SEM analysis of the materials revealed different microstructures even though they belong to the same class of materials. It could be shown that µ-CT and SEM imaging are valuable tools to understand microstructure and related mechanical properties of materials.

**Highlights**

- µ-CT and SEM imaging are valuable tools to understand material microstructure
- Mechanical properties might be derivable from the microstructure of the material
- Deviations from a homogeneous microstructure may influence material properties

1. Introduction

In recent years, various CAD/CAM (computer-aided design/computer-aided manufacturing) hybrid materials for permanent restorations appeared on the dental market [1]. They belong neither to polymers nor to ceramics but claim to combine the positive effects of ceramics and resin-based polymer materials [1]. The ranges of indication are similar and merge into each other, including the manufacturing of indirect restorations, such as inlays, onlays, partial and full crowns. CAD/CAM hybrid materials are characterized by organic and inorganic components [2]. Within the complete digital workflow these restorations can be produced by milling (subtractive manufacturing) or three-dimensional (3D) printing (additive manufacturing). Light-curable composites with inorganic components are also available as so-called nanoceramic composites for direct restorations [3]. The aim of all materials is to replace damaged and missing tooth structures.

Until recently CAD/CAM hybrid materials for permanent indirect restorations were mostly available as millable blocks. Within the last years a ceramic filled resin for 3D-printing by digital light processing (DLP) became clinically available (VarseoSmile Crown plus, BEGO, Bremen, Germany). It is approved for the application as adhesively cemented permanent restorations. DLP is the most widely applied 3D-printing technology for dental applications. This technology utilizes a digital light projection to crosslink photopolymerizable liquid resins layer by layer. DLP features more economical material consumption and an even more efficient digital workflow than subtractive manufacturing [4]. However, one requirement for materials to be usable for DLP is a suitable viscosity [5]. The filler content is decisive because it correlates with the viscosity positively. Therefore, the higher the filler content of a material is, the higher the viscosity of a material will be. In DLP resin filler content is therefore generally lower than in CAD/CAM millable hybrid material blocks to ensure a printable resin with suitable viscosity for processing in commercial dental DLP printers. The filler amount is in a similar range as in nanoceramic composites for direct restorations, in which the filler content is also limited due to a low viscosity required during intraoral modelling [6]. However, no studies are available regarding the microstructure and the filler distribution of CAD/CAM millable and printable hybrid materials for indirect restorations compared to nanoceramic composites for direct restorations.

It is unknown to what extent the materials differ in terms of their microstructure because they all belong to the same class of materials but differ in processing. An analysis of the microstructure can address various factors, including filler distribution, particle size and frequency and size of pores [7-9]. Variations in microstructure influence mechanical, chemical and biological properties. Consequently, the long-term stability of newly available CAD/CAM hybrid millable and printable materials is largely unknown compared to all-ceramic materials [6, 10-14]. A first preclinical analysis of CAD/CAM hybrid materials for milling and printing regarding the flexural strength and fatigue behaviour was conducted [15]. It could be shown that the CAD/CAM hybrid material for 3D-printing exhibited the lowest mechanical properties. A reason might be the inhomogeneous microstructure through the mixing procedure [15]. In addition, the size and distribution of the fillers may correlate with polishability and colour stability [16, 17]. Furthermore, the mechanical strength of a material is essential for its clinical use and longevity [17]. Particularly in the case of CAD/CAM printable hybrid materials, data is scarce. To the best of our knowledge, no statements can yet be made about their microstructure.

To evaluate the microstructure of CAD/CAM millable and printable hybrid materials, micro-computed tomography (μ-CT) analysis is a promising technique. Elliot et al. first introduced and established it as a dental research analysis [18]. Since then, the tool has become very popular due to its ability to visualize and analyse specimens in 3D in a non-destructive way [19, 20]. It has been used extensively in biomaterial studies, especially in the analysis of dental composite materials [19, 21-25]. Synchrotron μ-CT enables the assessment of the 3D microstructure of dental materials at both higher sensitivity and higher spatial resolution compared to conventional laboratory μ-CT [26-28].

In the present study, three CAD/CAM millable and printable hybrid materials of different compositions were analysed using synchrotron μ-CT and scanning electron microscopy (SEM) analysis. The CAD/CAM hybrid materials included one printable material for additive manufacturing by DLP (VarseoSmile Crown plus, BEGO, Bremen, Germany) and two CAD/CAM millable hybrid materials (Vita Enamic blocs, Vita Zahnfabrik, Bad Säckingen, Germany; Grandio, Voco, Cuxhaven, Germany). Additionally, one nanoceramic composite for direct restorations (Ceram X duo, Dentsply DeTrey, Konstanz, Germany) was analysed as a control group. The aim of this study was to assess and compare the material micro-morphology. Potential material properties might be derivable from two-dimensional (2D) and 3D images.

## 2. Material and Methods

The dental materials analysed in this study, including their composition properties published by the manufacturers, are listed in Table 1.

### 2.1 Geometry of the specimens

Cylindrical specimens of each of the CAD/CAM millable and printable hybrid materials were created using an STL dataset designed in FreeCAD (Version 0.19.2 for Windows) [29]. The diameter was 2 mm, and the length was 15 mm. The CAD/CAM millable and printable hybrid materials were produced and post-processed according to the manufacturer's instructions. A transparent cylindrical mould with the same dimensions was produced for the direct composite material. The nanoceramic composite for direct restorations was manually layered and light cured for 20 s (VALO Cordless, Ultradent Products, Cologne, Germany) from each side.

2.2 Data collection

The SEM images were collected on a Phenom XL electron microscope (ThermoFisher Scientific, Waltham, MA, USA). The samples were embedded, polished, and subsequently sputter coated with gold to avoid charging during SEM imaging. Imaging was conducted at 10 keV in backscattered mode (BSE) to show elemental contrast.

**Table 1:** Analysed materials.

| Material | Composition | Manufacturer | Code |
|---|---|---|---|
| **VarseoSmile Crown plus** | Ceramic-filled (30-50 wt% inorganic fillers; particle size 0.7 µm) silanized dental glass, methyl benzoylfor-mate, diphenyl (2,4, 6-trimethylbenzoyl) phosphine oxide hybrid material | BEGO, Bremen, Germany | VSCP |
| **Vita Enamic** | Polymer infiltrated (Urethane dimethacrylate, Triethylenglycoldimethacrylat 14 wt%) feldspar ceramic network (86 wt%) | VITA Zahnfabrik, Bad Sackingen, Germany | VE |
| **Voco Grandio** | Resin nanohybrid composite (86 wt% inorganic fillers; particle size 20-60 nm), embedded in a polymer matrix (14% UDMA + DMA) | Voco, Cuxhaven, Germany | VG |
| **Ceram.x duo** | Universal nano ceramic resin composite (glass filler content 76 wt%; particle size 10 nm), methacrylate modified | Dentsply DeTrey, Konstanz, Germany | CX |

Synchrotron X-ray µ-CT data were acquired at the ANATOMIX beamline of the French synchrotron light source SOLEIL (Saint-Aubin, France) [30] using a polychromatic beam with a central photon energy around 40-45 keV. Each tomogram consisted of a total of 4000 radiographs acquired over an angular range of 360°. The acquisition time was set to 200 ms per radiograph. The propagation distance between sample and detection plane was 50 mm. The detector consisted of a fluorescent screen (single-crystal lutetium aluminum garnet $Lu_3Al_5O_{12}$, 20 µm thick, supplier: Crytur, Turnov, Czech Republic), microscope optics with a 10×/0.28NA objective (Mitutoyo), and a CMOS-based camera (Orca Flash 4.0 V2, Hamamatsu Photonics K.K., Japan), resulting in an isotropic pixel size of 0.648 µm.

The tomography volume data were reconstructed from the projection radiographs with the PyHST2 [31] software (ESRF, Grenoble, France) and Paganin's method [32], in combination with the conventional filtered back-projection algorithm using a Paganin length of 65 pixels. The dimensions of the resulting reconstructed volumes were approximately 2.5 mm in diameter and 1.3 mm in height.

**2.3 Image data processing**

Data visualization was performed using ImageJ (distribution FIJI [33]) and Avizo (v. 2019.3; Thermo Fisher Scientific Inc., Waltham, MA, USA).

Both in-house developed Python tools and IPSDK (v. 3.1.0.1, Reactiv'IP, Grenoble, France) were applied to segment each image volume into its different material phases, i.e., the ceramic particles (or particle agglomerations) and air inclusions (if present). It was not always possible to distinguish between empty pores and pores filled with a weakly-absorbing material such as polymer or resin. Segmentation was performed using thresholding methods in combination with morphological operations. Due to the large number of small particles or particle agglomerations, a size filter with a radius of 3 pixels was applied using morphological operations (i.e. opening by reconstruction), which removed all detected strongly absorbing objects smaller than 3.9 µm in diameter. The masks of the segmented and filtered particles or particle agglomerates were then used as input for a connected component analysis to compute the particle size, shape and spatial orientation distribution for each of the four different samples, which were also composed of different components. The particle size is expressed in equivalent spherical diameter (ESD).

To study the impact of larger agglomerations within each material distribution, two further morphological size filters, with radii of 5 and 7 pixels, were applied to virtually remove all agglomerations with less than, respectively, 6.5 µm and 9.1 µm diameter (Figure 1).

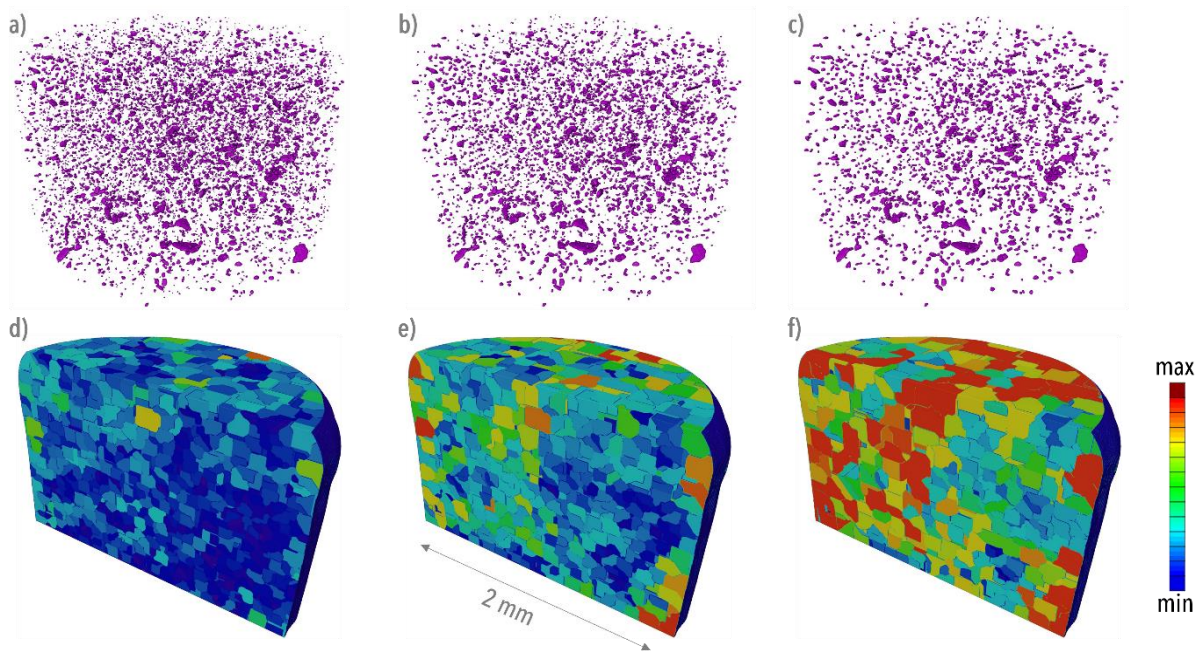

**Figure 1:** 3D rendering of segmented particles or particle agglomerations of one specific sample (VSCP) after opening by reconstruction with a radius of 3 px (a), 5 px (b) and 7 px (c) with corresponding basins after watershed segmentation (d-f). Blue indicates a smaller basin size and red a larger one. Note that, as the diameter of the morphological opening operation increases, small agglomerations are virtually removed so that the basin size of each remaining particle or individual particle agglomeration increases.

To quantify the 3D distribution of the particles or particle agglomerations, a watershed segmentation was applied to determine the volume surrounding each individual particle or particle agglomeration. All pixels around a particle that are closer to it than to any other particle are assigned to the particle's sphere of influence, or "basin". The size and variance of these basins are used to draw conclusions about the homogeneity of the spatial distribution of particles or particle agglomerations bigger than, respectively, 3.9 µm, 6.5 µm or 9.1 µm in diameter (Figure 1). A perfectly equal spatial distribution of particles or particle agglomerations would result in a very homogeneous basin-size distribution, whereas clustering of particles or particle agglomerations results in a broader basin-size distribution.

## 3. Results

### 3.1 Descriptive analysis of the tested materials

### 3.1.1 VarseoSmile Crown plus

The low-resolution scanning electron micrograph of the CAD/CAM printable hybrid material (VSCP) revealed horizontally oriented structures and some bigger white spots (Figure 2a). The nearly horizontal structures were spaced by around 50 µm. These layers were not completely straight and homogeneous and tend to sag in the center. At higher resolution, the microstructure and filler distribution resembled the CAD/CAM millable hybrid materials, with a homogenous distribution and very fine particle size below 1 µm. But focusing on the brighter, highly absorbing areas revealed that these represent agglomerates of various sizes of fine filler particles (inset Figure 2a). A homogeneous distribution of uniform fillers could therefore not be demonstrated (inset Figure 2a).

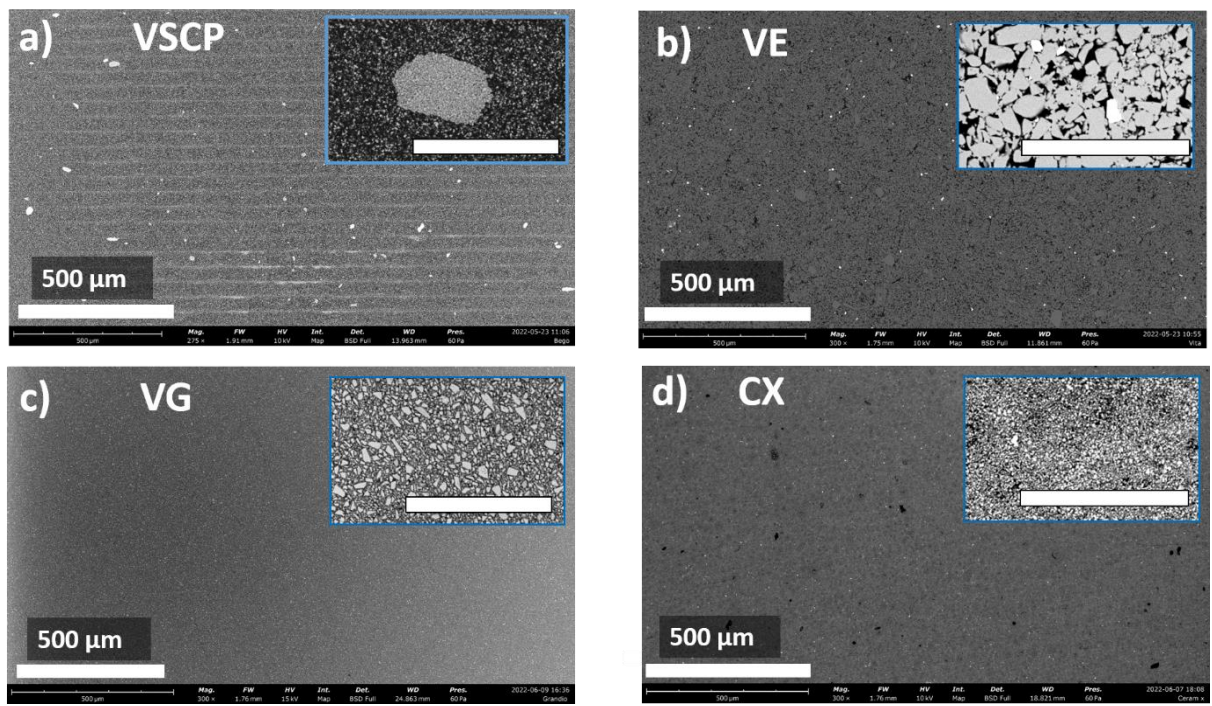

**Figure 2:** SEM images of the analysed materials: a) VSCP at low resolution. Inset shows one agglomerate in the VSCP at high resolution (inset scale bar length = 50 µm); b) SEM images of the material VE at low resolution with inset high resolution (inset bar length = 50 µm); c) SEM images of the material VG at low resolution with inset high resolution (inset bar length = 50 µm); d) SEM images of the material Ceram.x duo at low resolution with inset high resolution (inset bar length = 50 µm).

In the µ-CT investigations of the CAD/CAM printable hybrid material (VSCP), the layer structure was clearly visible throughout the investigated volume (Figure 3). Regular structures that ran almost parallel to each other were identifiable also at 50 µm intervals (Figure 3). Similar to the low-resolution SEM image, bright, highly absorbing regions were also visible in the volume investigation. As heavy elements are more radiopaque than light elements, the bright spots probably correspond to highly absorbing heavy particles, e.g., ceramic particles or agglomerations. In addition, individually occurring spherical dark areas were recognized which can be attributed to air bubbles (Figure 3c and d). One SEM image and a virtual cross section through a µ-CT image of VSCP in similar resolution were investigated for gray values. Corresponding line profiles of the gray values from both SEM and µ-CT data are shown in Figure 4. They are very similar, with a periodic change from lighter to darker gray values at around 50 µm. This corresponds well with the printing layers in the VSCP material, which were set to 50 µm. This reproducible modulation of gray value indicates a variation in material density within each layer (Figure 4).

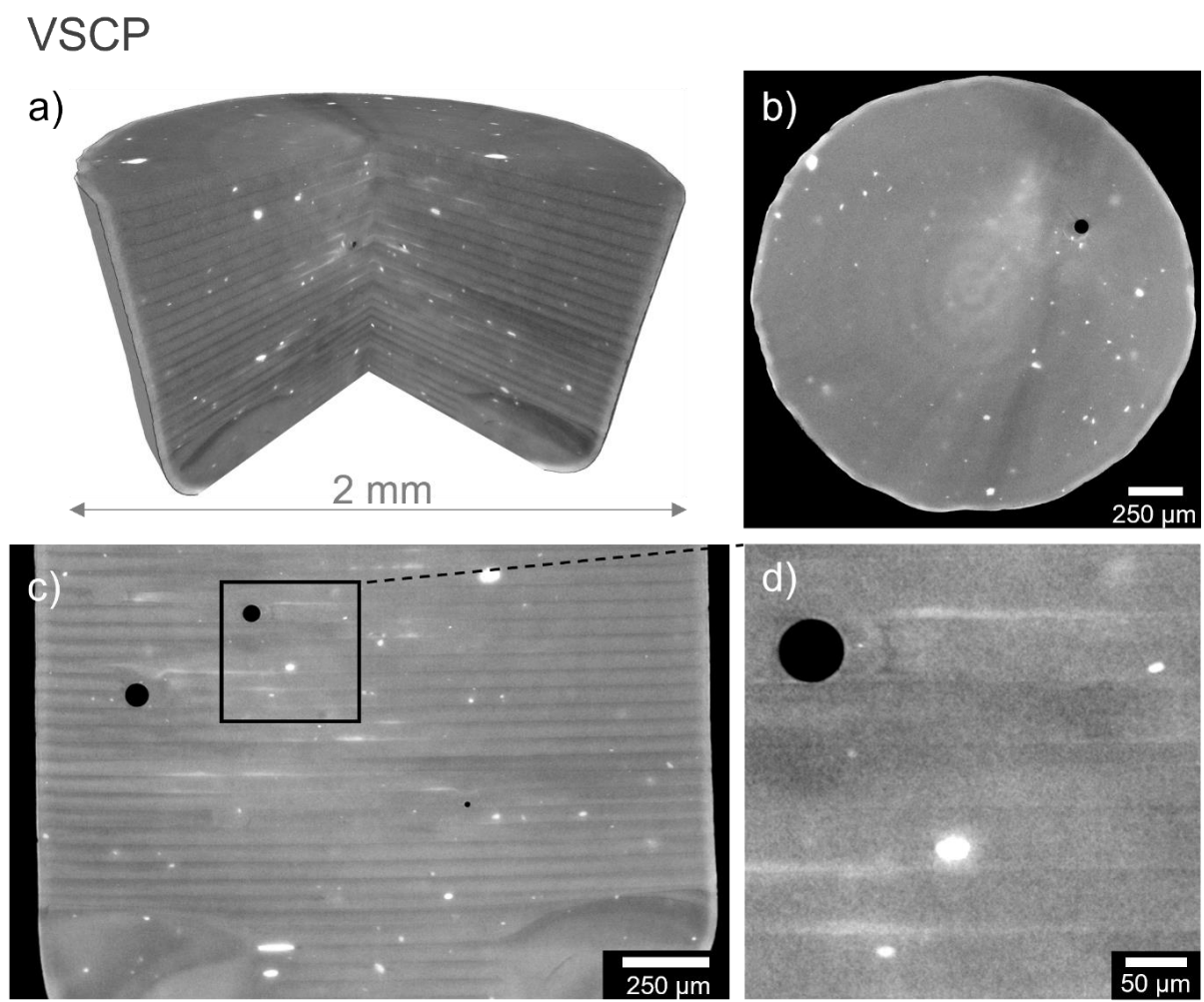

Figure 3: Micro-CT data of VSCP. a) Volume rendering. b,c) Virtual slices in b) horizontal and c) vertical plane, and d) enlarged detail of vertical slice. Layer structure as well as bright spots and dark areas are clearly visible.

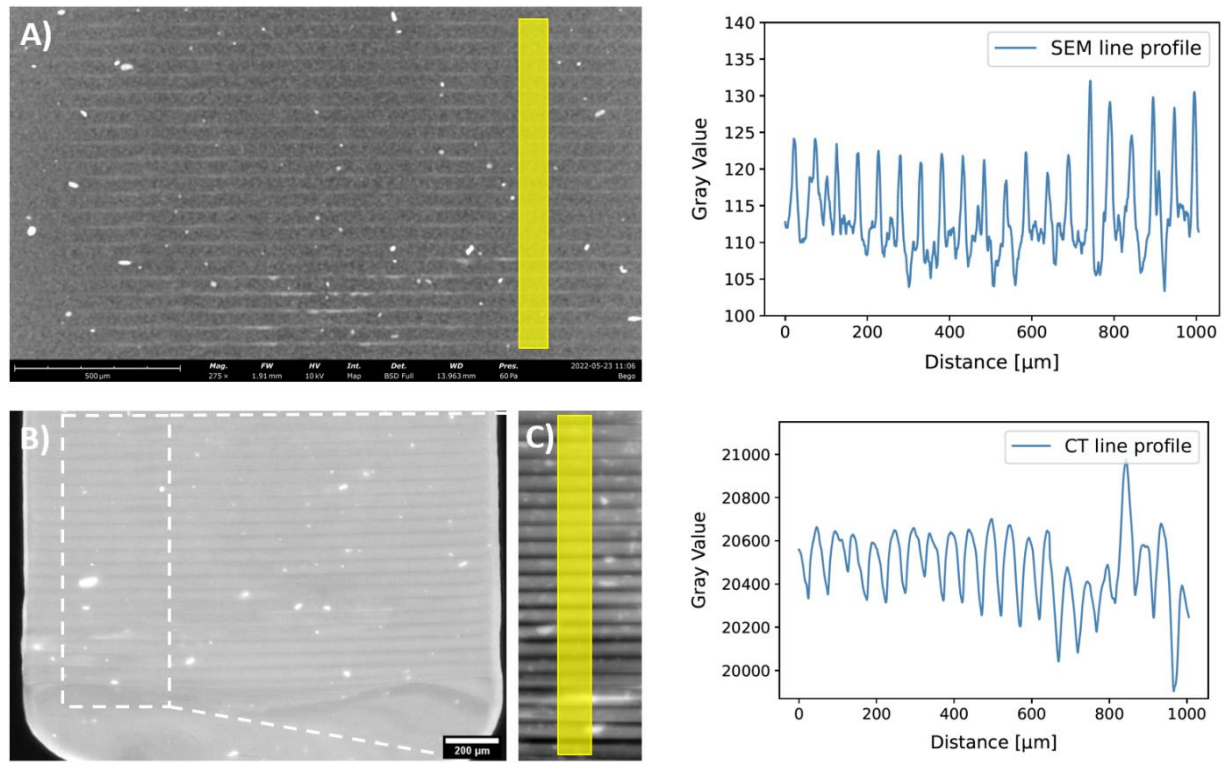

**Figure 4:** Filtered SEM image of VSCP at low resolution with yellow marker perpendicular to the 3D printed layers (A). Corresponding line profile (top right). Virtual cross section through the CT volume of VSCP (B) and section with integrated gray values along 1800 slices of the CT volume with yellow marker perpendicular to the 3D printed layers (C). Corresponding line profile (bottom right).

Additional analysis of the orientation of the particles or agglomerations of particles of VSCP revealed pronounced directional alignments within one plane, namely parallel to the layered structures associated to the printing plane (Figure 5 A). This preferred orientation is visible through maxima in the angles θ and ρ at 0 and 180°, which are both parallel to the x-y plane. The distribution of the ψ angle is relatively equal (Figure 5 E), indicating no preferred orientation of these particles or agglomerates within the x-y plane.

Most of these particles or agglomerations are elongated and arranged within the assumed printing layers, and show an aspect ratio between 0.35 and 0.5. This preferred orientation is visible through maxima in the theta angles at 0 and 180°, which is perpendicular to z.

### 3.1.2 Vita Enamic

The low-resolution scanning electron micrograph of Vita Enamic (VE) shows a uniform appearance with some variations in gray scale, indicating larger ceramic particles surrounded by an organic phase (Figure 2b). At higher resolution, the polymer-infiltrated ceramic network became visible with larger ceramic filler particles with a size up to 10 µm surrounded by the polymer network (inset Figure 2b). In the X-ray tomography images (shown in supplemental material, Figure S1), larger particles with heavier elemental composition are visible in a lighter gray scale due to higher absorbance, and the organic polymer matrix is visible against an almost black background due to light elements and lower absorption. Small and more homogeneously distributed opaque particles are visible (Figure S1).

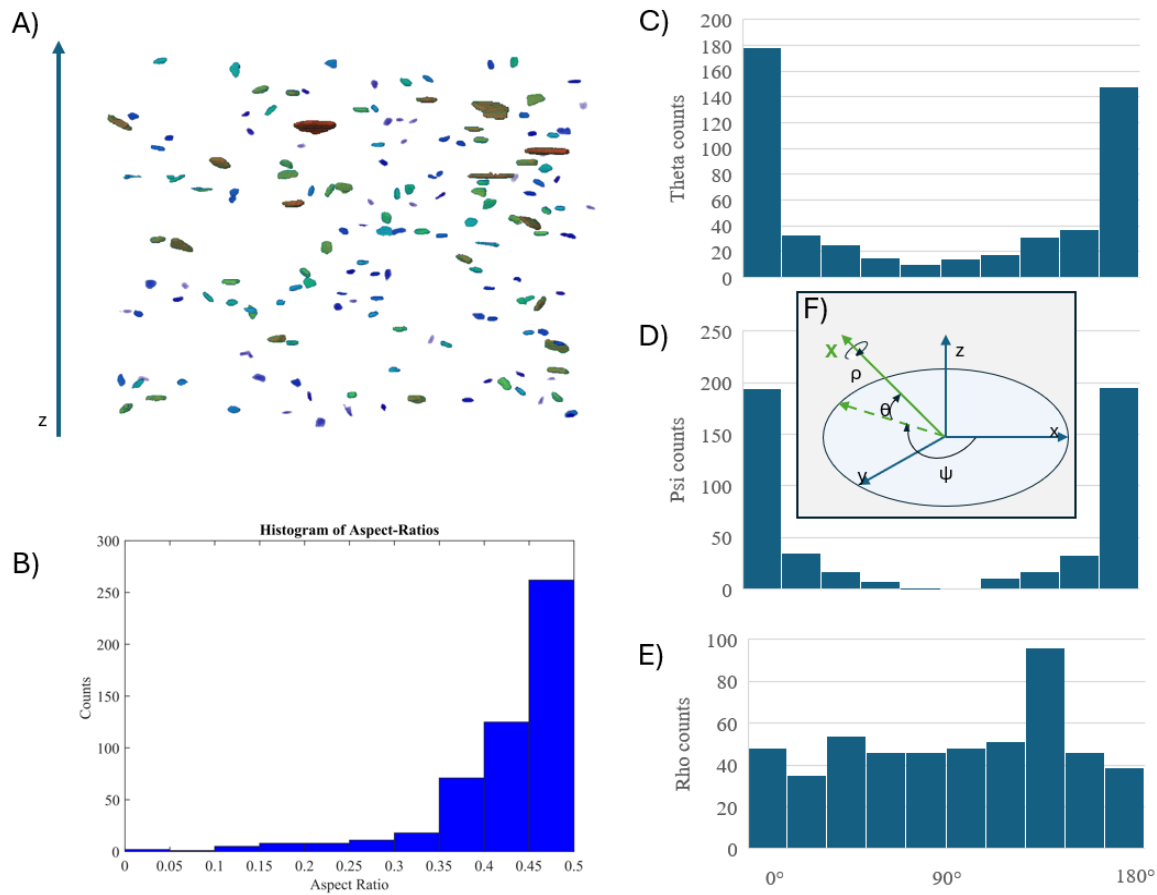

**Figure 5:** A) Colour coded (blue = 6.5 µm equivalent diameter, red = 290 µm equivalent diameter) 3D-rendering of the particles or particle agglomerations with an aspect ratio smaller than 0.5 and an equivalent diameter higher than 6.5 µm, B) histogram of the aspect ratios, C, D, E) histograms of the particle orientations *θ, ψ* and *ρ* and their respective orientations within the coordinate system (F).

### 3.1.3 Voco Grandio

The low-resolution scanning electron micrographs for Voco Grandio (VG) showed a homogeneous distribution of the fillers with no visible pores or agglomerations (Figure 2c). In high resolution (inset Figure 2c), uniform small ceramic fillers with particle sizes up to 1 µm are visible. The µ-CT data show that the material components of VG are distributed very homogeneously. Some minor irregularities can be seen, possibly pores (Figure S2).

### 3.1.4 Ceram.x duo

The scanning electron micrographs of the direct composite material (CX) at low resolution showed a homogenous microstructure with visible dark spots that can be attributed to pores (Figure 2d). At higher resolution, the uniform distribution of the filler particles is visible (inset Figure 2d). The µ-CT data of CX revealed both distinct translucent areas (potentially pores) and dense inclusions (Figure S3).

## 3.2 Microstructural comparison of the analysed materials

### 3.2.1 Filler distribution

The µ-CT-based comparison with respect to particle distribution with particles greater than 3.9 µm in diameter for all analysed materials revealed many large, irregularly arranged particles or particle agglomerations in VSCP, compared to the other investigated materials (Figure 6). The irregular filler distribution in VSCP was also visible (Figure 6). VE exhibited a uniform filler distribution. However, this filler size was bigger than in VG (Figure 6), which contains small and homogeneously distributed fillers (Figure 6). CX showed various small fillers greater than 3.9 µm in diameter (Figure 6).

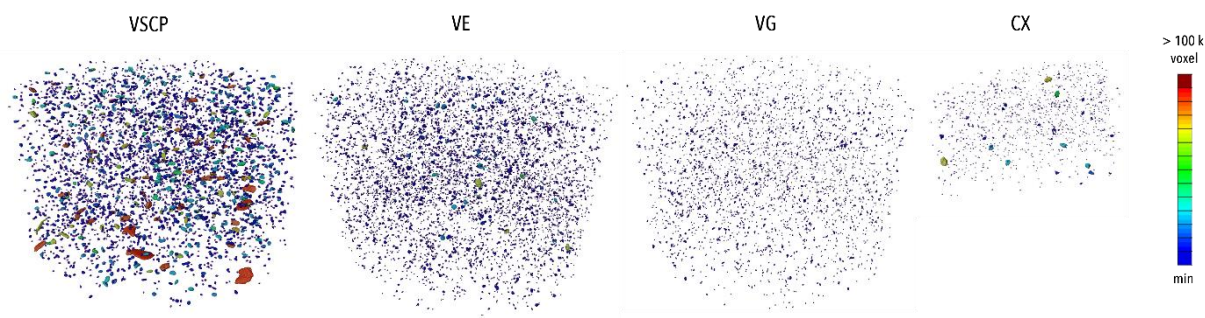

**Figure 6:** 3D rendering of the inclusions (colour code: small inclusions = blue, bigger inclusions = red) from µ-CT data of each sample after an opening by reconstruction with a spherical structuring element of radius 3 px (ø 3.9 µm). Note that only 1/8 of the volume of sample CX could be analysed due to the presence of image artefacts in the rest of the volume.

### 3.2.2 Particle size and frequency of pores

A comparative analysis of the materials in terms of their distribution of dense inclusions was performed through so-called basins (Figure 7). With each "opening" operation, particles below a certain radius were virtually removed from the analysis of each material. It could thus be shown that the number of particles in VSCP did not change much depending on the different "openings". At an opening of 7 px radius (corresponding to ø 9.1 µm particle or particle agglomeration size) VSCP still contained a lot of particles (Figure 7 d–f for VSCP), meaning that a lot of detected particles or particle agglomerations in this material were larger than 9.1 µm. By comparison, VE showed a substantial decrease of the number of particles as the opening radius was increased from 3 to 7 px. Overall, it showed slightly larger particles or particle agglomerations than VG, a finding supported by SEM investigations. When an opening of 5 px radius (corresponding to ø 6.5 µm) was applied, more particles or particle agglomerations were visible in VE than in the other CAD/CAM millable hybrid material, VG. VG showed much lower particle or particle agglomeration content at this level of image opening, as the particle or particle agglomerations size of this material was in the range of 3 px (ø 3.9 µm) (Figure 7). The small particle or particle agglomerations described for CX at an opening of 3 px (ø 3.9 µm) were no longer visible at even higher values (5 and 7 px) for the image opening. Only porosities in the sense of air bubbles were visible here (Figure 7).

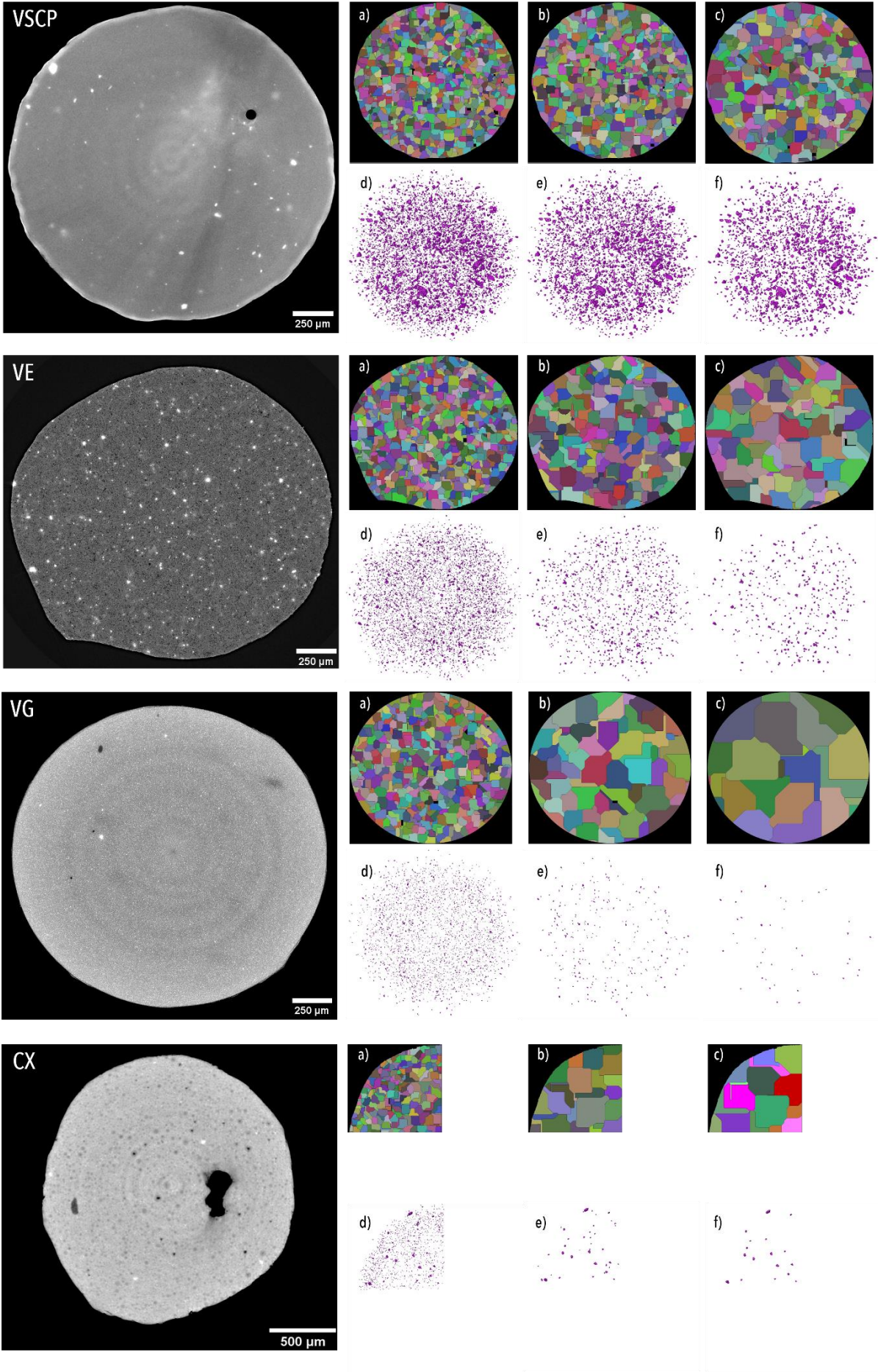

**Figure 7** (preceding page): Virtual μ-CT slice in xy plane for each material (VSCP, VE, VG, CX) and illustration of the basins after morphological opening by reconstruction on the segmented inclusions with radii of 3 px (a), 5 px (b) and 7 px (c) for each material as well a 3D rendering of the corresponding inclusions.

Error! Reference source not found. shows the quantification of the porosities and particles after the removal of smaller particle or particle agglomerations. The overall porosity was highest in the Ceram.x duo sample (0.82 % of the total investigated volume) and lowest in VE (0; no pores) and VG (0; no pores). VSCP revealed a mean overall porosity of 0.02 % of the total investigated volume. Regarding the quantification of the particles the volume of particles in the VSCP material is only slightly reduced with each increase in opening radius, from 0.0141 mm³ (particle agglomerations > 3.9 μm Ø) to 0.0115 mm³ (particle agglomerations > 9.1 μm Ø). This shows that most detected particles or particle agglomerations in the VSCP material were larger than 9.1 μm Ø. They make up for around 0.3 % of the total investigated volume. The volume of inclusions larger than 3.9 μm in the VE material was 0.0033 mm³. The volume decreases with an increase in opening radius to 0.002 mm³ for particle agglomerations > 6.5 μm Ø and to 0.0012 mm³ for particle agglomerations > 9.1 μm Ø. This shows that there are different sized particles or particle agglomerations, and only 0.032 % of the total investigated volume is accounted for by particles or particle agglomerations larger than 9.1 μm. For the VG material the total volume of inclusions was even smaller, ranging from 0.001 mm³ for particles or particle agglomerations > 3.9 μm Ø to 0.0001 mm³ for particles or particle agglomerations > 9.1 μm Ø. This was supported by the SEM investigation, which showed particles with a small and uniform particle size below 5 μm in the VG material. The CX material showed the smallest volume of inclusions, ranging from 0.0003 mm³ for particles or particle agglomerations > 3.9 μm Ø to 0.0001 mm³ for particles or particle agglomerations > 9.1 μm Ø. This also aligns with the SEM investigation, which showed that most particles were around or below 1 μm in size.

**Table 2:** Volume in mm³ and percentage of total volume (TV) of the segmented inclusions for all analysed materials after an opening by reconstruction with 3 px, 5 px or 7 px radii, respectively. Note the big differences between the % of the TV of VSCP and all other materials, especially for inclusions bigger than 9.1 μm in diameter.

| Material | Total volume [mm³] | Particle agglomerations | | | | | | Porosity | |
|---|---|---|---|---|---|---|---|---|---|
| | | > 3.9 μm Ø | | > 6.5 μm Ø | | > 9.1 μm Ø | | | |
| | | [mm³] | % of TV | [mm³] | % of TV | [mm³] | % of TV | [mm³] | % of TV |
| VSCP | 4.09 | 0.0141 | 0.345 | 0.0132 | 0.322 | 0.0115 | 0.281 | 0.000712 | 0.0174 |
| VE | 3.84 | 0.0033 | 0.086 | 0.002 | 0.053 | 0.0012 | 0.032 | 0 | 0 |
| VG | 3.76 | 0.001 | 0.027 | 0.0003 | 0.0076 | 0.0001 | 0.0022 | 0 | 0 |
| CX | 0.38 | 0.0003 | 0.075 | 0.00015 | 0.04 | 0.0001 | 0.035 | 0.003 | 0.819 |

The means and the standard deviations (SDs) of the basin volumes are listed in Table 3. Basins are employed to quantify the spatial distribution of the particles. Large mean values of the basins correspond to fewer particles (> than 3.9 or 6.5 or 9.1 μm) and smaller SD values indicate a more homogenous spatial distribution of these particles. The ratio of SD over mean is computed to compare the homogeneity of the basin-size distribution between the four different samples after virtually removing particles with a diameter smaller than 3.9 µm, 6.5 µm and 9.1 µm. Thus, a higher SD-to-mean ratio indicates higher heterogeneity of spatial particle distribution. The SD-to-mean ratio of the VSCP sample was much less dependent on the considered particle sizes than that of the other four samples. The highest impact of virtual removal of smaller particles on the spatial distribution was found for sample CX, indicating a higher degree of heterogeneity (Figure 7).

**Table 3:** Mean, standard deviation (SD) and SD/mean of the volume of the basins obtained through the watershed segmentation. Note that the SD/mean is within the same range for all samples.

| Material | VSCP | | |
|---|---|---|---|
| particle agglomerations | 3.9 µm | 6.5 µm | 9.1 µm |
| Mean [µm³] | $4.62 \times 10^5$ | $8.36 \times 10^5$ | $1.5 \times 10^6$ |
| SD [µm³] | $2.61 \times 10^5$ | $4.54 \times 10^5$ | $8.31 \times 10^5$ |
| SD / mean | 0.56 | 0.54 | 0.54 |
| Material | VE | | |
| opening diameter | 3.9 µm | 6.5 µm | 9.1 µm |
| Mean [µm³] | $5.51 \times 10^5$ | $2.71 \times 10^6$ | $9.31 \times 10^6$ |
| SD [µm³] | $2.77 \times 10^5$ | $1.43 \times 10^6$ | $5.26 \times 10^6$ |
| SD / mean | 0.50 | 0.53 | 0.56 |
| Material | VG | | |
| opening diameter | 3.9 µm | 6.5 µm | 9.1 µm |
| Mean [µm³] | $9.71 \times 10^5$ | $1.35 \times 10^7$ | $8.13 \times 10^7$ |
| SD [µm³] | $4.59 \times 10^5$ | $7.19 \times 10^6$ | $5.35 \times 10^7$ |
| SD / mean | 0.47 | 0.53 | 0.66 |
| Material | CX | | |
| opening diameter | 3.9 µm | 6.5 µm | 9.1 µm |
| Mean [µm³] | $3.13 \times 10^5$ | $8.40 \times 10^6$ | $2.00 \times 10^7$ |
| SD [µm³] | $2.20 \times 10^5$ | $4.88 \times 10^6$ | $9.99 \times 10^6$ |
| SD / mean | 0.70 | 0.58 | 0.50 |

The size distribution of the dense particles and their basins is also shown in Figure 8, revealing the narrowest size distribution of the dense particles for CX and the widest size distribution for VSCP. By far, VSCP has the highest number of larger particles, as revealed by the green histograms (Figure 8). The shape of the basins' histograms, in turn, indicates strong differences for the spatial distribution of the dense particles between the four samples towards higher basin volumes in sample VE (Figure 8). Considering only basins formed by the larger particles (yellow and green), one can see that only VSCP and VE have a sufficient number of large particles to compare the corresponding basins (Figure 8).

Information on size and shape of the particle agglomerations with diameters exceeding 9.1 µm for the samples VSCP, VE, VG, and CX are listed in Table 4. Smaller particles are excluded from the shape evaluation, as they would strongly distort the values. Each sample is characterized by several parameters including the total number of inclusions, the size of inclusions and equivalent diameter (mean, standard deviation, and maximum), aspect ratio and sphericity (mean and standard deviation). Only a few particles larger than 9.1 µm equivalent diameter (ED) remain for the samples CX and VG. Sample CX has the largest remaining particles with a mean of 17.4 µm ED and sample VG the smallest with a mean of 11.4 µm ED. The aspect ratio and the sphericity values are very similar for all four samples.

**Table 4:** Volume, equivalent diameter, aspect ratio and sphericity of the analysed particles obtained through the watershed segmentation and a closing by reconstruction of 7 pixel. Note that for samples CX and VG only a few particles or particle agglomerations remain.

|  | Particle agglomerations > 9.1 µm Ø | | | |
|---|---|---|---|---|
| **Sample** | **VSCP** | **VE** | **VG** | **CX** |
| **total number of inclusions:** | **2515** | **367** | **46** | **17** |
| **Inclusion Size [µm³]** | | | | |
| mean: | 4578 | 3373 | 1816 | 7823 |
| std: | 9756 | 3123 | 1036 | 6825 |
| max: | 271883 | 22900 | 7909 | 22918 |
| **Equivalent Diameter [µm]** | | | | |
| mean: | 14.1 | 13.5 | 11.4 | 17.4 |
| std: | 4.6 | 3.2 | 1.6 | 5.3 |
| max: | 61.8 | 27.1 | 19.0 | 27.1 |
| **Aspect Ratio** | | | | |
| mean: | 0.65 | 0.69 | 0.64 | 0.66 |
| std: | 0.12 | 0.10 | 0.09 | 0.11 |
| **Sphericity** | | | | |
| mean: | 0.77 | 0.72 | 0.74 | 0.71 |
| std: | 0.06 | 0.08 | 0.05 | 0.08 |

The direct comparison of SEM with µ-CT imaging is shown in supplementary figure S4. It can be seen that mass-density contrast is very high for µ-CT but also that the image resolution of the SEM is much better resulting in sharper images at the shown resolution (Figure S4). The selected SEM and µ-CT images show similarities in microstructure. Inorganic fillers are clearly visible due to absorption contrast in both SEM and µ-CT (Figure S4). SEM only allows for 2D imaging in selected areas, but can give higher resolution than µ-CT, while µ-CT allows investigations of a complete volume, providing data with statistical significance. But this always strongly depends on the resolution, and in case of hybrid materials with sub-micrometer inorganic fillers this tool can only be used to investigate the overall distribution of structures larger than (in our case) 3.9 µm. This limitation affects inorganic fillers and possible porosity investigations as well.

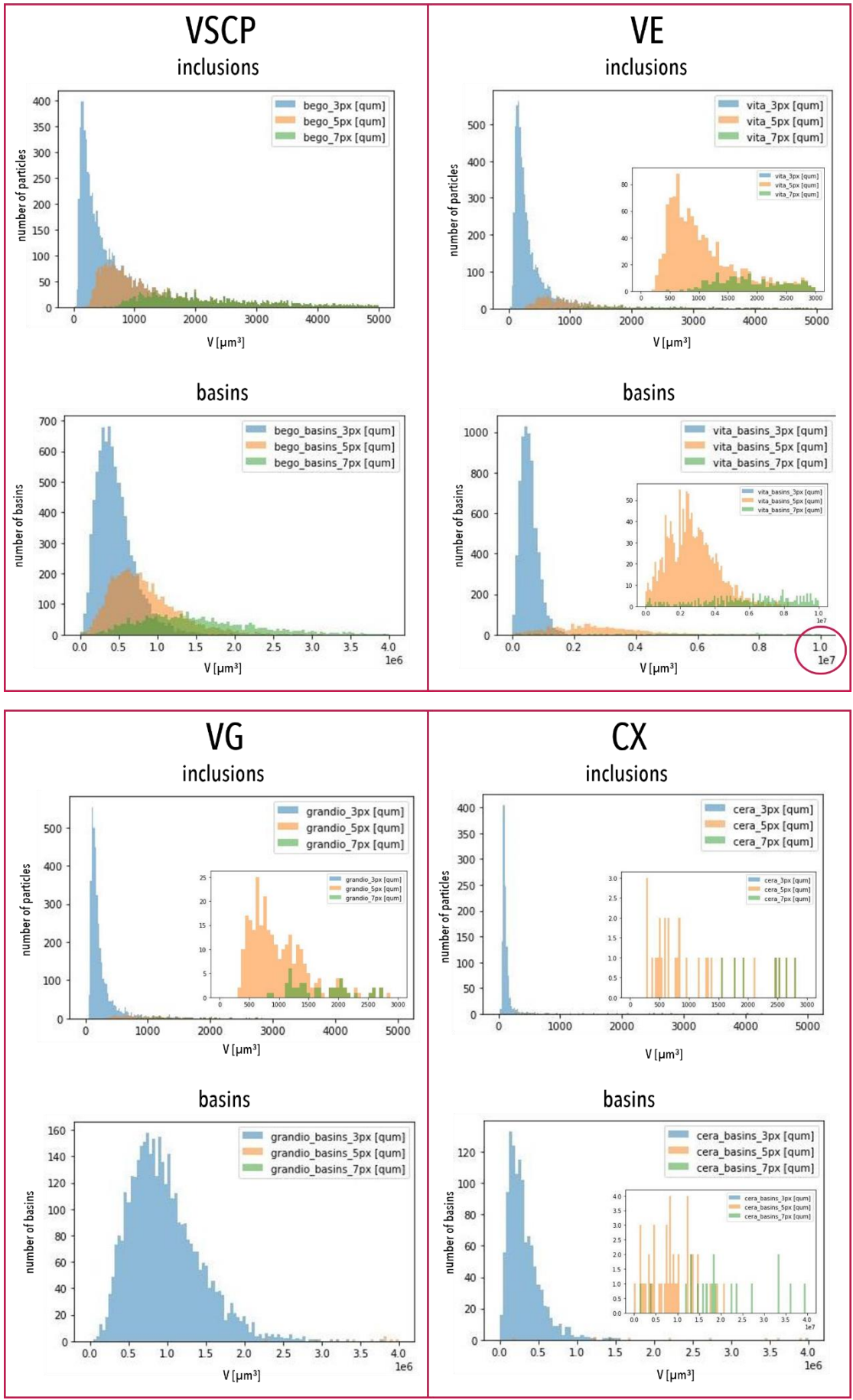

**Figure 8** (preceding page): Histograms of the size distribution of the inclusions and their basins are shown after virtually removing inclusions with a radius smaller than 3px (blue), 5px (orange) and 7 px (green).

## 4. Discussion

The aim of this study was to present an analysis of the microstructure of different CAD/CAM millable hybrid materials (VE and VG), one CAD/CAM hybrid printable material (VSCP) and one direct composite material (CX) based on µ-CT and SEM images, as these materials have similar indications. The µ-CT analysis showed clear differences in the structural composition of the investigated materials.

Before the printing process of the CAD/CAM printable hybrid material (VSCP) starts, a homogeneous distribution of the ceramic fillers can be supported by different stabilizing chemicals, but cannot be guaranteed, as they can sink or segregate in the liquid resin over time. However, the filler content and its distribution also play a decisive role for the subsequent material-specific properties of the 3D-printed restorations. The filler content of the CAD/CAM 3D-printable hybrid materials is often lower than that of the CAD/CAM millable hybrid materials available in blocks to allow free flow of the material in the vat of the DLP printer prior to layerwise photopolymerization [34]. A low filler content in the final 3D-printed component in addition affects the stiffness of the material and results in an elastic modulus that is much lower than that of natural hard dental tissues [34-36]. A previous study already described a random distribution of the inorganic filler but did not discuss any agglomerates [34]. It was also concluded that the flexural modulus was significantly reduced due to a low filler content in comparison to millable materials of similar composition [34]. Rodriguez et al. showed that an increase in filler content may lead to an increase in compressive strength and flexural modulus, but the relationship strongly depends on the filler particle size and size distribution [37]. They found that designed nanoparticle agglomerates increased compressive strength more than just well distributed nanoparticles [37]. In our study the investigated VSCP material showed a distribution of fine particles, as proven by SEM images, in addition to larger agglomerations of these particles, with up to 62 µm equivalent diameter (Table 4). According to Rodriguez et al. this may enhance compressive strength compared to a uniform nano-sized filler distribution [37]. Due to the printing process most of these larger agglomerates were found to be oriented parallel to the printing orientation (Figure 5). Depending on the printing direction of indirect restorations this preferred orientation could potentially enhance the compressive and flexural strength of the produced crown [38]. Furthermore, it cannot be excluded that polishability and colour stability are also influenced by an irregular composition of the material. The printing direction could also have an influence on the wear resistance and colour stability of the restorations [34]. According to Grzebieluch et al., the 3D-printed restorations should be placed in such a way that the tensile force during mastication is applied along and not across the layers [34]. However, literature on 3D printing in dentistry is sparse so far [34]. Future in vitro studies could complement the present study. The µ-CT data suggest that it may be advantageous to carefully shake the printable material before printing, to ensure proper mixing of the components. Furthermore, the printing structure could be shown in SEM images for the first time. In addition, heterogeneous microstructures due to the printing layers were seen when comparing SEM and µ-CT images. The printing direction may affect material properties and therefore seems to be important for clinical application. The different microstructures of the tested materials could be important regarding mechanical properties. It could already been shown that CAD/CAM hybrid materials for milling and printing showed different microstructures and therefore different mechanical properties [15]. A lower initial biaxial flexural strength could be proven for VSCP [15]. Besides a lower biaxial flexural fatigue strength of VSCP could be shown. Consequently, a inhomogeneous microstructure of CAD/CAM hybrid materials for 3D printing might favor fracture [15].

In the present study, the industrially produced, CAD/CAM millable hybrid materials (VE and VG) seem to exhibit a homogeneous distribution of their constituents. Due to industrial production, high material homogeneity can be ensured compared to direct composite materials [6, 39, 40]. CAD/CAM millable hybrid blocks are divided according to their microstructure into dispersed fillers, such as VG, and polymer-infiltrated ceramic networks, such as VE [6, 39]. Dispersed fillers are characterized by a high proportion of ceramic fillers in a resin-based matrix [6]. In the case of VE the polymer-infiltrated ceramic networks consist of a pre-sintered glass-ceramic network that is infiltrated with monomers [6]. The 3D ceramic scaffold allows the monomers to interconnect [6]. This is the essential difference to dispersed fillers. Stresses can be effectively absorbed. Fractures or chippings can be avoided [6, 14]. Moreover, industrial processes allow an increase of the filler content [6]. A high filler content increases the mechanical stability, hardness, and elasticity modulus of a material [6, 41-43]. Wear resistance is considered a multiparametric property of materials [6]. It strongly depends on the filler content [6, 41, 44] as well as on the particle size, geometry and distribution [6, 45] of the fillers. According to a clinical study, high filler content and small particle sizes are considered advantageous [6, 46]. Based on the results of the present study, the mechanical stability of VE (polymer infiltrated ceramic network) and VG (nano-hybrid composite) might be superior to the other materials investigated because of the high filler content and low porosity detected in the µ-CT and SEM investigation. The polishability and colour stability of VG could be promising due to the high and homogeneous filler content and small particle size.

The direct composite material (CX) showed clear inhomogeneities, which represent multiple large air bubbles. The necessity of manual modelling and the microstructure of direct composite materials causes this specific microradiographic appearance. Porosities lead to reduced mechanical stability and wear resistance. Furthermore, poorer colour stability can be assumed. Nevertheless, direct composite materials are considered the material of choice for the restoration of class I and II cavities in posterior teeth [6, 47] because indirect restorations would result in further loss of dental hard tissues due to a preparation of the cavity design [6].

The µ-CT comparison of all analysed materials showed that the CAD/CAM printable hybrid material (VSCP) can be placed between the CAD/CAM millable hybrid materials (VE and VG) and the direct composite material (CX) in terms of filler distribution and porosity, as it showed lower porosity than CX, but also around 10x more large particle agglomerations > 9.1 µm than this material and all millable materials VE and VG. Therefore, it is to be expected that the mechanical stability of CAD/CAM hybrid materials for milling and printing is better than that of direct composite materials. So far, preclinical studies regarding the material properties of 3D-printed restorations are scarce. Additive technology is considered to have a high development potential in dentistry. CAD/CAM millable hybrid materials, on the other hand, are established in a fully digital workflow. They showed good mechanical results in vitro [6].

In the present study, only one sample of each material was tested. Future investigation should include more samples to minimize the variations induced by the manufacturing of the sample. Due to the limited resolution of the µ-CT imaging configuration used, only inclusions (particles and particle agglomerations as well as pores) with a diameter larger than 3.9 µm were included in the µ-CT-based analysis. From the SEM images, however, it becomes obvious that there is a large number of much smaller particles that also vary between the different samples. Although the role of nano-sized particles for the mechanical stability was discussed in the introduction, they cannot be assessed here. Thus, ignoring these small particles will have a strong impact on our results and the derived conclusions. This shortcoming is compensated by including additional SEM images into this study. However, SEM only provides 2D information and thus only probes a small fraction of the sample as compared to 3D µ-CT imaging. Finally, emerging X-ray nanotomography methods can give access to shorter length scales than "classical" synchrotron µ-CT, albeit at smaller probed volumes [48].

## 5. Conclusions

The CAD/CAM printable hybrid material showed a more heterogeneous distribution of the filler particles than the other CAD/CAM millable hybrid material but a more homogeneous structure than the direct composite material. The CAD/CAM millable hybrid materials showed a very homogeneous arrangement of the structures, which can be attributed to the industrial production. The direct composite material displayed considerable inhomogeneities, particularly air pockets due to manual plugging. The CAD/CAM printable hybrid material showed a layer structure caused by the printing process. Information on the microstructural composition of the analysed materials could allow researchers to draw conclusions about additional material properties which could have an impact on the materials' clinical suitability. In the future, various 3D-printable CAD/CAM hybrid materials could be compared. According to the µ-CT evaluation of the present study, it can be assumed that the CAD/CAM printable hybrid material can be located between the CAD/CAM millable hybrid materials and the direct composite material regarding filler distribution.

## 6. Acknowledgments

### 6.1 Funding

This work was supported by the Deutsche Gesellschaft für Prothetische Zahnmedizin und Biomaterialien (DGPro) [no grant number given by the DGPro].

ANATOMIX is an Equipment of Excellence (EQUIPEX) funded by the *Investments for the Future* programme of the French National Research Agency (ANR), project *NanoimagesX*, grant no. ANR-11-EQPX-0031.

### 6.2 Declarations of interest

None

Supplementary data

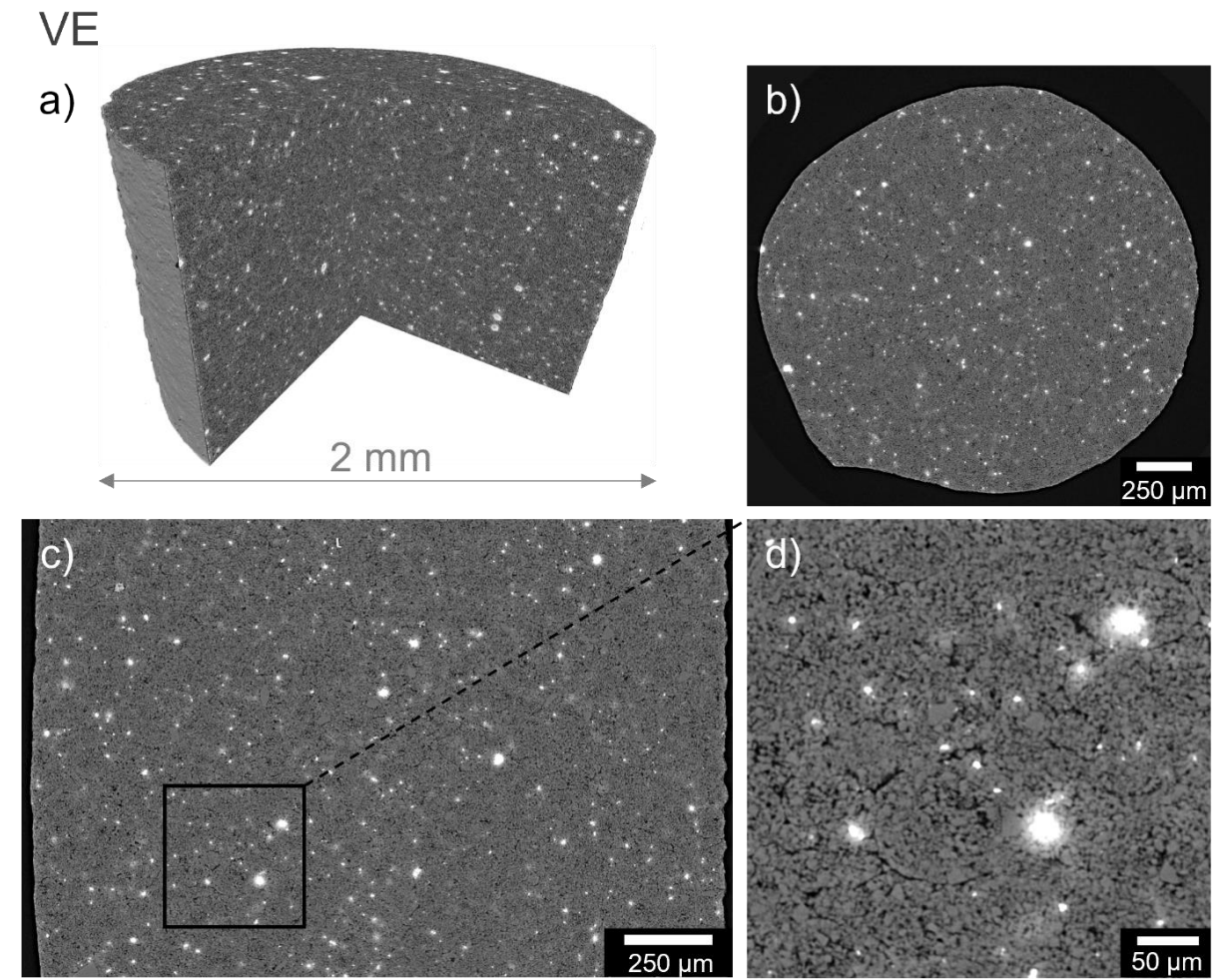

**Figure S1:** a) Volume rendering for VE and b) virtual slices in horizontal and c) vertical-plane with d) detail view.

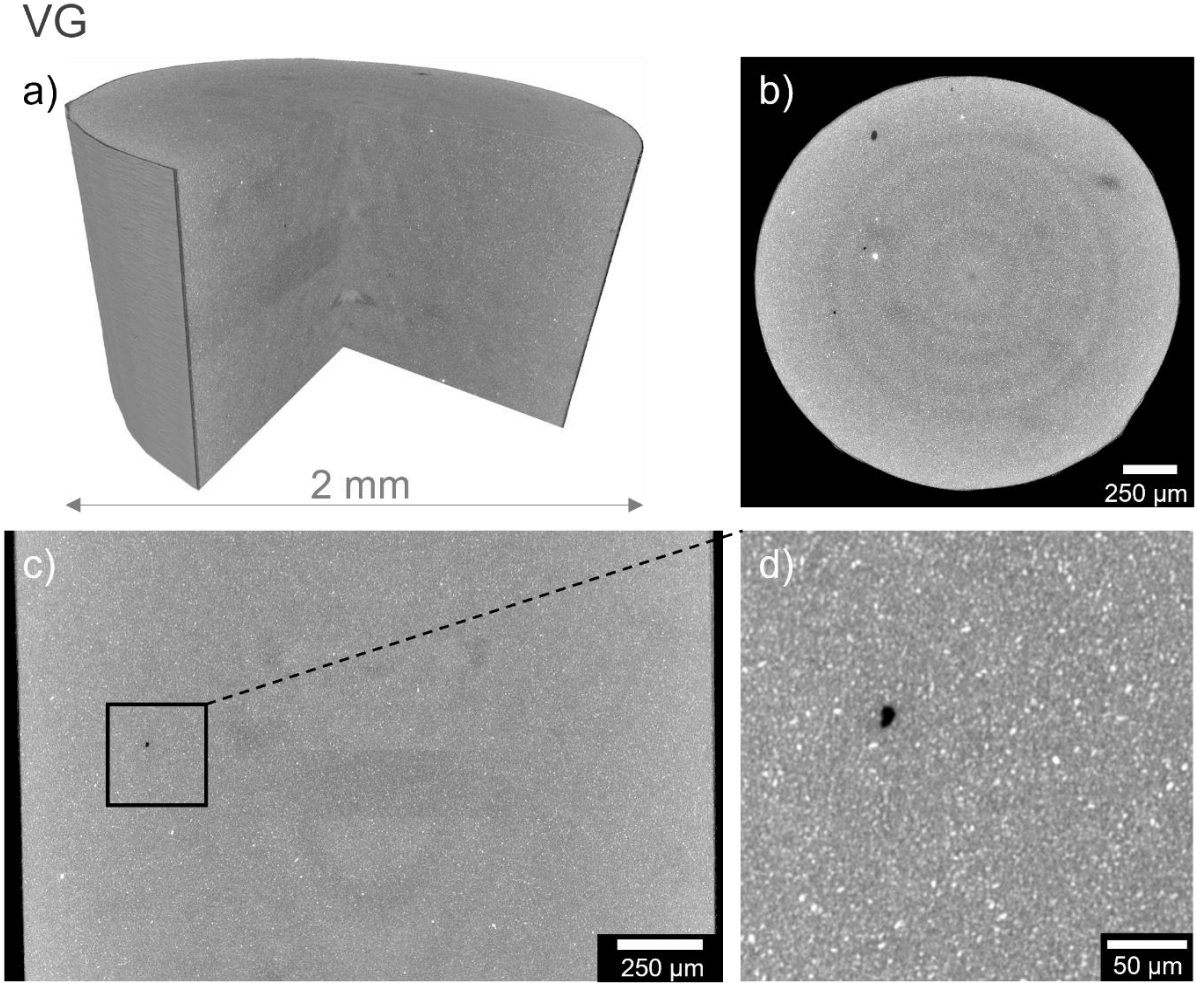

**Figure S2:** a) Volume rendering for VG and b) virtual slices in horizontal and c) vertical-plane with d) detail view.

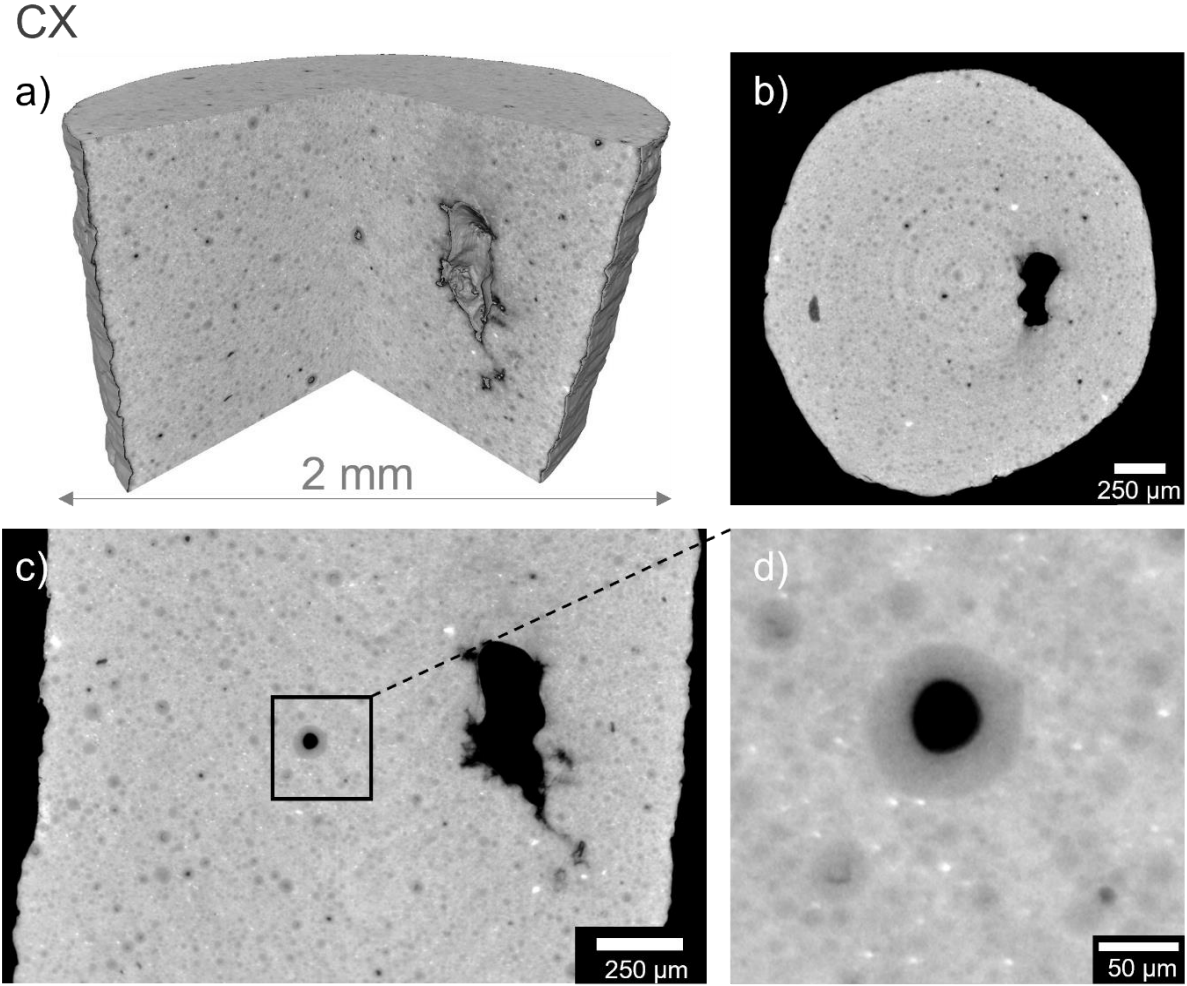

**Figure S3:** a) Volume rendering for CX and b) virtual slices in horizontal and c) vertical-plane with d) detail view.

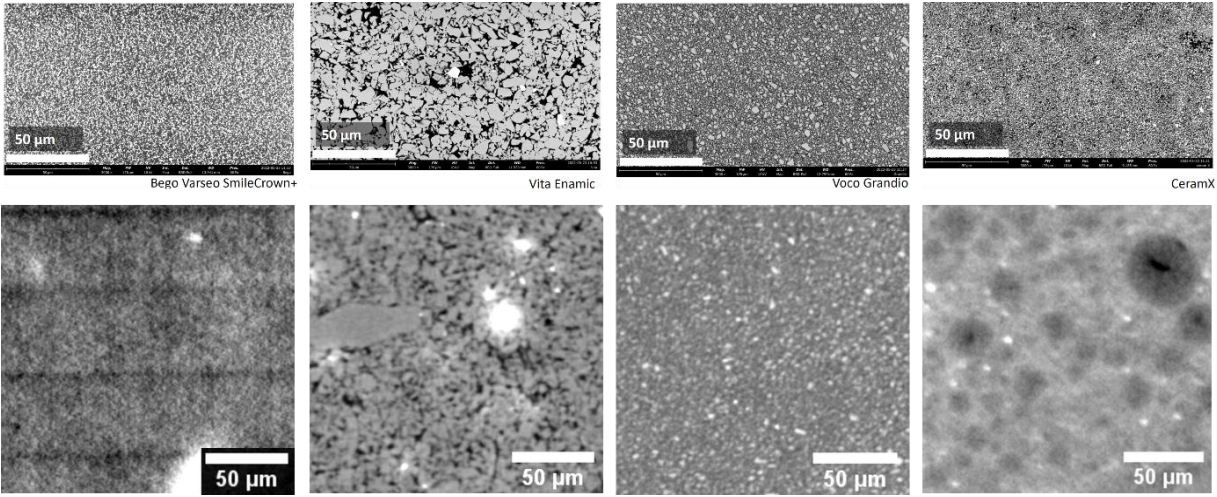

**Figure S4:** Comparison of both modalities SEM (top) and µCT (bottom) for all four materials VSCP, VE, VG and CX from left to right.